\documentclass{emulateapj}
\usepackage{apjfonts}
 
\usepackage{bm}
\usepackage{color}
\usepackage{graphicx}
\shorttitle{Spontaneous Formation of Surface Magnetic Structure from Strongly-stratified Large-scale Dynamo}
\shortauthors{Masada and Sano}
\begin{document}
\title{Spontaneous Formation of Surface Magnetic Structure \\from Large-scale Dynamo in Strongly-stratified Convection}
\author{Youhei MASADA\altaffilmark{1}, and Takayoshi SANO\altaffilmark{2}} 
\altaffiltext{1}{Department of Physics and Astronomy, Aichi University of Education; Kariya, Aichi 446-8501, Japan: 
E-mail: ymasada@auecc.aichi-edu.ac.jp}
\altaffiltext{2}{Institute of Laser Engineering, Osaka University; Osaka 565-0871, Japan: E-mail: sano@ile.osaka-u.ac.jp}
\begin{abstract}
We report the first successful simulation of spontaneous formation of surface magnetic structures from a large-scale dynamo by
strongly-stratified thermal convection in Cartesian geometry. The large-scale dynamo observed in our strongly-stratified model has
physical properties similar to those in earlier weakly-stratified convective dynamo simulations, indicating that the $\alpha^2$-type
mechanism is responsible for it. Additionally to the large-scale dynamo, we find that large-scale structures of the vertical magnetic 
field are spontaneously formed in the convection zone surface only for the case of strongly-stratified atmosphere. The organization 
of the vertical magnetic field proceeds in the upper convection zone within tens of convective turn-over time and band-like bipolar 
structures are recurrently-appeared in the dynamo-saturated stage. We examine possibilities of several candidates as the 
origin of the surface magnetic structure formation, and then suggest the existence of an as-yet-unknown mechanism for the 
self-organization of the large-scale magnetic structure, which should be inherent in the strongly-stratified convective atmosphere. 
\end{abstract}
\section{Introduction} 
A long-standing goal of the solar interior physics is to reproduce active regions, composed mainly of the sunspots, self-consistently from 
magnetic fluxes generated in the solar interior. We now approach the subject from two different theoretical perspectives: one focuses on 
emergence and organization processes of the magnetic flux in the uppermost part of the convection zone (CZ), and the other explores 
the flux generation and maintenance processes, i.e., the dynamo process, operating deeper down.  

Several leading-edge numerical studies, which focus on the uppermost part of the solar CZ, have succeeded to simulate spontaneous 
formations of concentrated magnetic structures reminiscent of active regions  \citep[e.g.,][]{cheung+10,stein+12,rempel+14,kapyla+15}. 
In these studies, the solar surface convection and its nonlinear interaction with the magnetic field were simulated in a more or less realistic 
manner with the steep density gradient just below the photosphere and/or the radiative transfer with the ionization in Cartesian domains. 
However, since some sort of the large-scale seed magnetic field has been assumed inconsistently as initial or boundary condition, the dynamo 
mechanism and its connection to the formation process of the active region were beyond the scope of these studies. 

A growing body of evidence is accumulating to demonstrate that solar-like cyclic large-scale magnetic field is organized in global 
spherical-shell convections \citep[e.g.,][]{ghizaru+10,kapyla+12,masada+13,yadav+15,augustson+15}. Despite some differences in the 
numerical setup and method, there is a common outcome of the convective dynamo in these studies: diffuse magnetic flux extending over 
the CZ and/or the tachocline instead of magnetic flux tubes expected in the standard solar dynamo paradigm \citep[e.g.,][and references 
therein]{charbonneau10}. Although the flux emergence like event from distributed magnetic flux has been occasionally observed in some 
models \citep[][]{nelson+13,fan+14}, its universality or feasibility in the Sun is still a matter of considerable debate. 

There is still a large gap between the dynamo in the interior and the active region formation at the surface. Our study in this Letter would 
be a first step aiming to bridge the gap between them. By advancing our previous works of weakly-stratified MHD convection 
\citep[][hereafter MS14a,b]{masada+14a,masada+14b}, we perform convective dynamo simulation in a strongly-stratified atmosphere 
resembling the solar interior in Cartesian geometry. The spontaneous formation of large-scale magnetic structures in the CZ surface 
self-consistently from the large-scale convective dynamo is reported. 
\section{Numerical setup}
Convective dynamo system is solved numerically in Cartesian domain. Our model covers only the CZ of depth $d_{\rm cz}$ 
($0 \le z \le d_{\rm cz}$) with omitting a stably-stratified layer below it, where $x$- and $y$-axes are taken to be horizontal and 
$z$-axis is pointing downward. We set the width of the domain to be $W = 4d_{\rm cz}$. 

We solve the fully-compressible MHD equations in the rotating frame of reference with a constant angular velocity of
${\bm \Omega} = -\Omega_0 {\bm e}_z$,
\begin{eqnarray}
\frac{\partial \rho}{\partial t} & = & - \nabla\cdot (\rho {\bm u})  \;, \label{eq1} \\ 
\frac{{\mathcal{D} }{\bm u}}{\mathcal{D} t}  & = & - \frac{\nabla P}{\rho}  
+ \frac{{\bm J}\times{\bm B} }{\rho} - 2{\bm \Omega}\times{\bm u} + \frac{\nabla \cdot {\bm \Pi}}{\rho} + {\bm g}  \;, \ \ \ \ \label{eq2} \\
\frac{\mathcal{D}\epsilon }{\mathcal{D} t} & = & - \frac{P\nabla\cdot {\bm u}}{\rho}  + \mathcal{Q}_{\rm heat} \;, \label{eq3} \\
\frac{\partial {\bm B} }{\partial t} & = & \nabla \times ({\bm u} \times {\bm B} - \eta_0 {\bm J})\;, \label{eq4}
\end{eqnarray}
with the viscous stress ${\bm \Pi}$ and the heating term $\mathcal{Q}_{\rm heat}$ of 
\begin{eqnarray}
 \Pi_{ij} &=& 2\rho\nu_0S_{ij} = \rho\nu_0\left( \frac{\partial u_i}{\partial x_j} +  \frac{\partial u_j}{\partial x_i} - \frac{2}{3}\delta_{ij}\frac{\partial u_i}{\partial x_i} \right) \;,  \label{eq5} \\
\mathcal{Q}_{\rm heat} &=&   \frac{\gamma\nabla\cdot(\kappa_0 \nabla\epsilon)}{\rho} + 2\nu_0 \bm{S}^2 + \frac{\mu_0\eta_0\bm{J}^2}{\rho}\;, \label{eq6}
\end{eqnarray}
where $\mathcal{D}/\mathcal{D}t$ is the total derivative, $\epsilon = c_{\rm V} T$ is the specific internal energy, 
${\bm J} = \nabla \times {\bm B}/\mu_0$ is the current density with the vacuum permeability $\mu_0$, and ${\bm g} = g_0 {\bm e}_z$ is the 
gravity of the constant $g_0$. The viscosity, magnetic diffusivity, and thermal conductivity are represented by $\nu_0$, $\eta_0$, and 
$\kappa_0$, respectively. A perfect gas law $P = (\gamma -1)\rho \epsilon$ with $\gamma = 5/3$ is assumed. 

The initial hydrostatic state is polytropic stratification given by 
\begin{equation}
\epsilon = \epsilon_0 + \frac{g_0z}{(\gamma-1)(m + 1)}\;, \ \ \ {\rm and} \ \ \ \rho = \rho_0 (\epsilon/\epsilon_0)^m \;,  
\end{equation}
with the initial surface internal energy $\epsilon_0$, surface density $\rho_0$, and the polytropic index $m = 1.49$, 
providing the super-adiabaticity of $\delta\ \equiv \nabla - \nabla_{\rm ad}\ = 1.6 \times 10^{-3}$, where 
$\nabla_{\rm ad} = 1-1/\gamma$ and $\nabla = (\partial \ln T/\partial \ln P)$.

Normalization quantities are defined by setting $d_{\rm cz}/2 = g_0= \rho_0 = \mu_0 = c_p = 1$. 
The normalized pressure scale-height at the surface, defined by $\xi = H_p/d_{\rm cz} = (\gamma-1)\epsilon_0/(g_0d_{\rm cz})$, controls 
the stratification level and is chosen here as $\xi = 0.01$, yielding a strong stratification with the density contrast between top and 
bottom CZs about $700$. 

Figure 1a shows the initial profiles of the density (solid) and temperature (dashed) of our model. The 
density profile in the range $0.71 \le r/R_\odot \le 0.991$ of the standard solar model is also shown in 
such a way as to fit the computational domain (dash-dotted) \citep[Model S:][]{CD+96}.  
Our model has a stratification almost encompassing the solar CZ except its uppermost part. 

All the variables are assumed to be horizontally periodic. Stress-free boundary conditions are used in the vertical 
direction for the velocity. Perfect conductor and vertical field conditions are used for the magnetic field at the bottom and top 
boundaries. A constant energy flux which drives the thermal convection is imposed on the bottom boundary, while the specific 
internal energy is fixed at the top boundary. 

The fundamental equations are solved by the second-order Godunov-type finite-difference scheme that employs an approximate MHD 
Riemann solver \citep{sano+98}. The magnetic field evolves with the Consistent MoC-CT method \citep{evans+88,clarke96}. 
Non-dimensional parameters ${\rm Pr} =20$, ${\rm Pm} =2$, ${\rm Ra} = 6\times10^7$, angular velocity of $\Omega_0 = 0.5$ and 
the spatial resolution of ($N_x, N_y, N_z$) $=$ $(256,256,256)$  are adopted, where the Prandtl, magnetic Prandtl, and Rayleigh 
numbers are defined by 
\begin{equation}
{\rm Pr}  =  \frac{\nu_0}{(\kappa_0/\rho c_p )} \;,\ {\rm Pm} = \frac{\nu_0}{\eta_0}\;, \ {\rm Ra}  =  \frac{g_0 d_{\rm cz}^4}{\chi_0 \nu_0}\frac{\delta}{H_{p}}  \;, 
\end{equation}
where $\rho$, $\delta$, and $H_p$ are evaluated at $z = d_{\rm cz}/2$.

In the following, the volume-, horizontal-, $x$- and $y$- averages are denoted by single angular brackets with subscripts ``\rm v", ``\rm h", ``\rm x"  
and ``\rm y", respectively. The time-average of each spatial mean is denoted by additional angular brackets. The 
relative importance of convection to the rotation is measured by the Rossby number ${\rm Ro} = (u_{\rm cv}k_f)/(2\Omega_0)$, where 
$u_{\rm cv} \equiv \sqrt{\langle \langle u_z^2 \rangle_{\rm v} \rangle}$ is the mean convective velocity, and $k_f = 2\pi/d_{\rm cz}$.  
The global convective turn-over time and the equipartition field strength are defined by 
$\tau_{\rm cv} \equiv 1/(u_{\rm cv}k_f)$ and $B_{\rm eq} (z) \equiv \sqrt{\langle \mu_0 \rho {\bm u}^2 \rangle_{\rm h}} $. 
Note that $B_{\rm eq}(z)$ is evaluated from the local convective energy and thus has a depth-dependence.

Since the sound speed in the deep CZ becomes very large in the strongly-stratified model and imposes a strict limit on the time-step,  
a long thermal relaxation time is required in our fully-compressible simulation. To alleviate it, we first construct a progenitor model, in 
which the convection reaches a fully-developed state and the system becomes thermally-relaxed, by evolving a non-rotating 
hydrodynamic run for $800\tau_{\rm cv}$. 

Shown in Figure 1b and 1c are the vertical profile of $\langle\langle u_z^2 \rangle\rangle_{\rm h}^{1/2}$ (dashed) and the distributions 
of $u_z$ in the $x$-$y$ plane at $z/d_{\rm cz} = 0.04,\ 0.23,\ 0.78$ in the equilibrated state of the progenitor model. The black (orange) 
tone denotes down-(up-)flows. The multi-scale convection with the strong up-down asymmetry, i.e., the slower and broader upflow cell 
surrounded by networks of the faster and narrower downflow lanes, is developed in the progenitor model \citep[e.g.,][]{spruit+90,miesch05}. The 
dynamo run is started by adding the rotation and a seed weak horizontal field to the progenitor model. 
\begin{figure}[tbp]
\begin{center}
\begin{tabular}{ccc}
\includegraphics[width=8.5cm,clip]{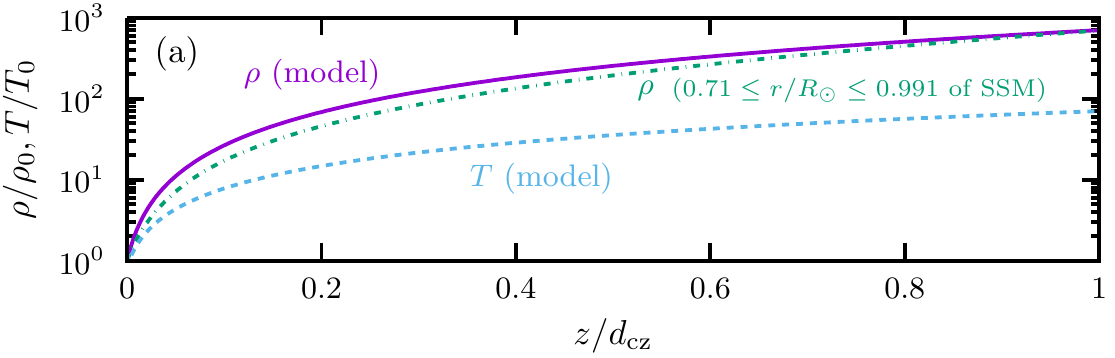} \\
\includegraphics[width=8.5cm,clip]{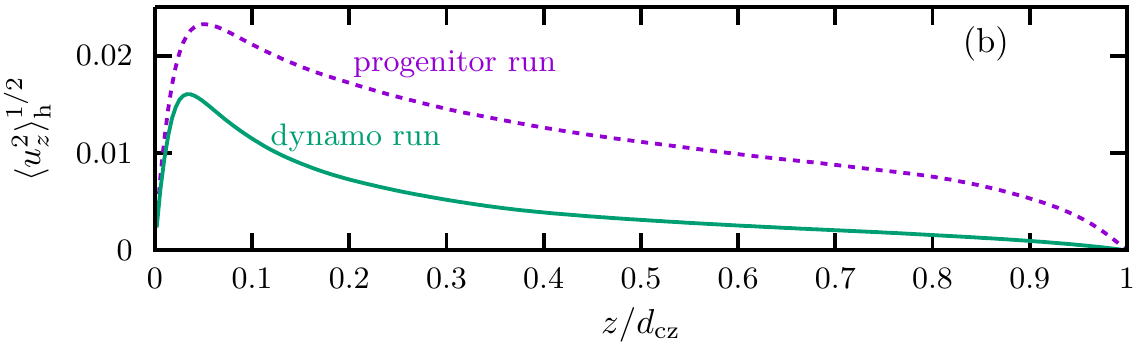} \\
\includegraphics[width=8.5cm,clip]{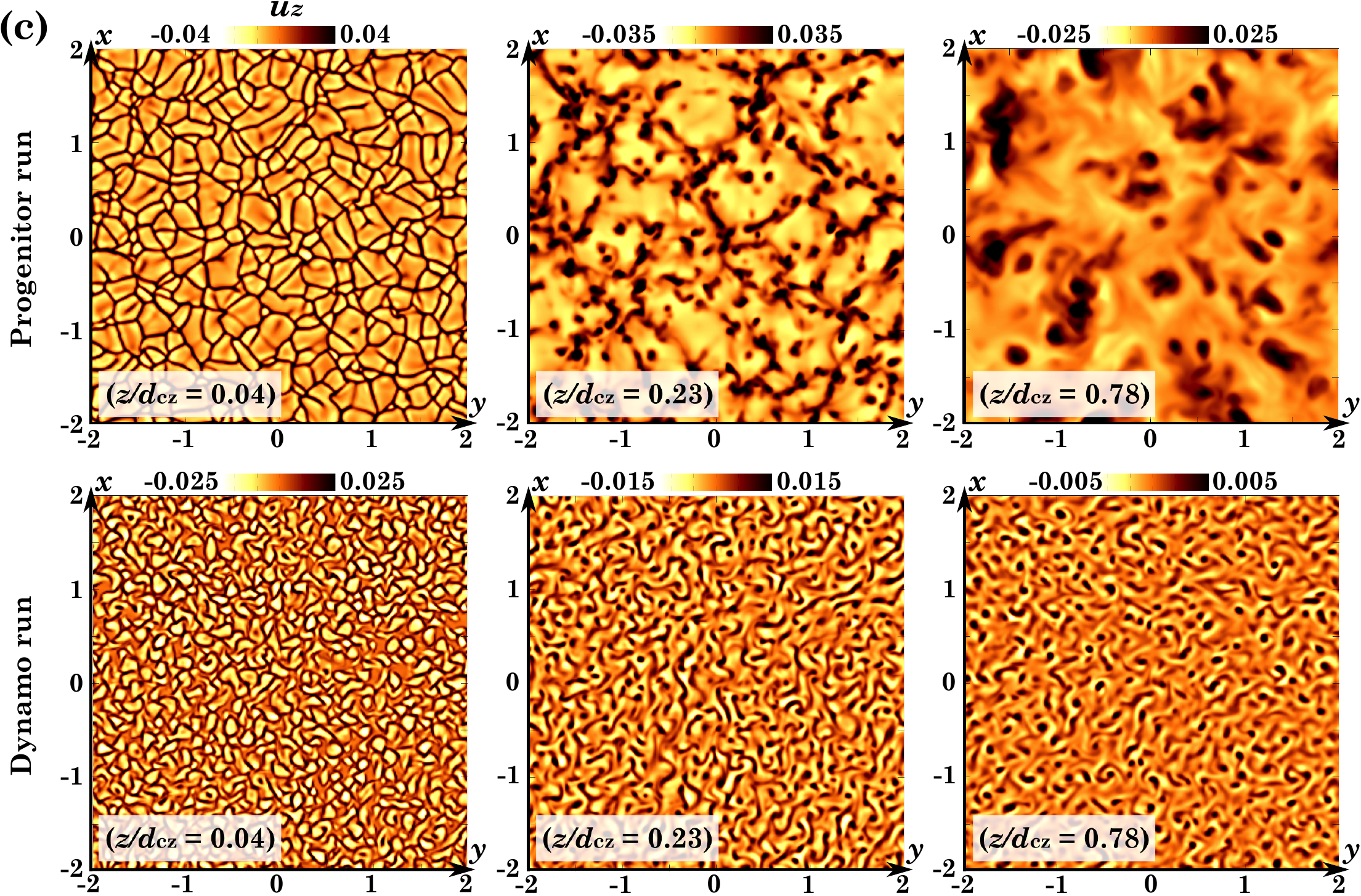}
\end{tabular}
\caption{(a) Vertical profiles of the initial density (solid) and temperature (dashed) of the simulation model, and the density profile of the 
CZ of the standard solar model (dash-dotted). Normalization units are their surface values. (b) Vertical profiles of $\langle\langle u_z^2 
\rangle\rangle_{\rm h}^{1/2}$ for the progenitor (dashed) and dynamo (solid) runs. (c) The horizontal distributions of the $u_z$ at 
$z/d_{\rm cz} = 0.04,\ 0.23,\ 0.78$ of the progenitor (upper) and dynamo (lower) runs. } 
\label{fig1}
\end{center}
\end{figure}
\begin{figure*}[htbp]
\begin{minipage}{0.35\hsize}
\begin{center}
\includegraphics[width=5.5cm,clip]{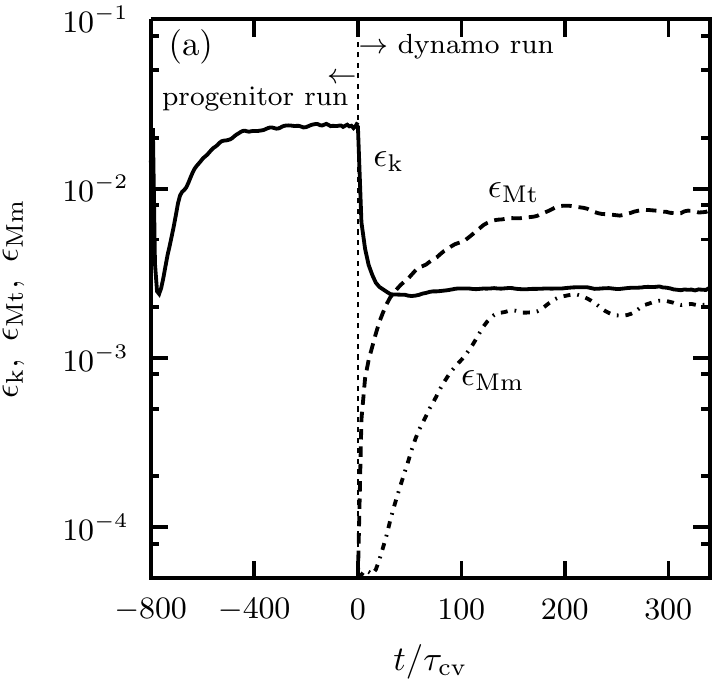}
\end{center}
\end{minipage}
\begin{minipage}{0.35\hsize}
\begin{center}
\includegraphics[width=10.5cm,clip]{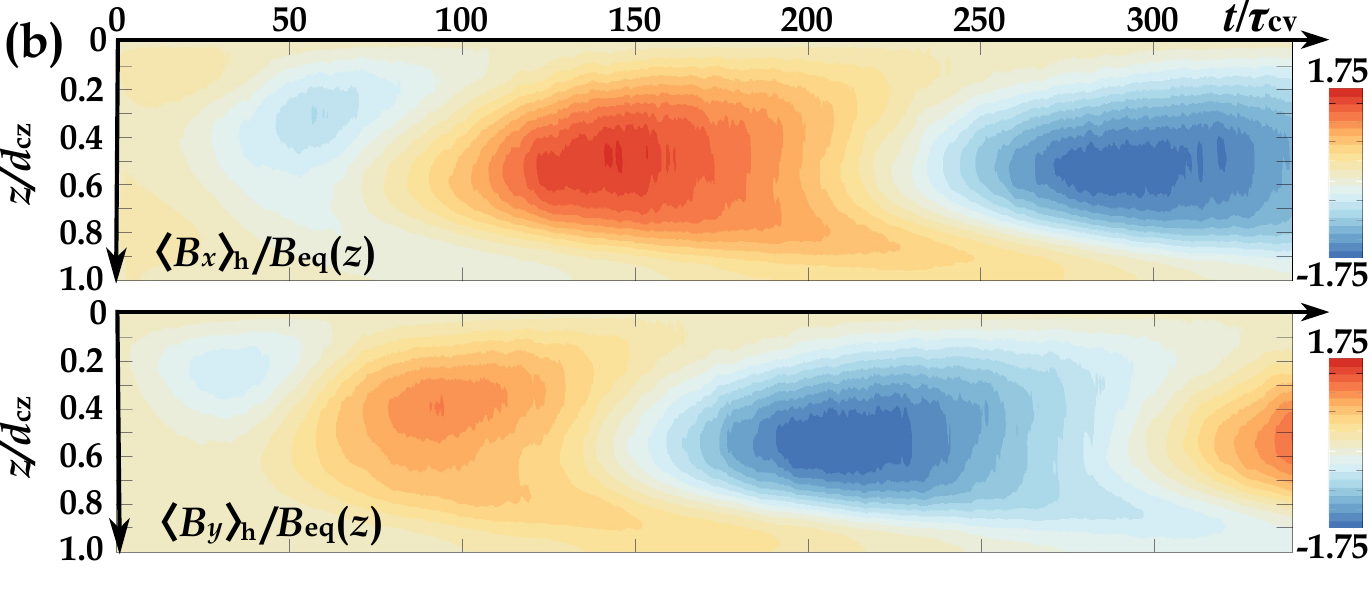}
\end{center}
\end{minipage}
\caption{(a) The temporal evolution of $\epsilon_{\rm K}$ (solid), $\epsilon_{\rm Mt}$ (dashed) and $\epsilon_{\rm Mm}$ (dash-dotted). 
(b) Time-depth diagrams of $\langle B_x \rangle_{\rm h}/B_{\rm eq}(z)$ and $\langle B_y \rangle_{\rm h}/B_{\rm eq}(z)$. } 
\label{fig2}
\end{figure*}
\section{Simulation Results}
\subsection{Basic Properties of Convection and Dynamo}
The temporal evolutions of $\epsilon_{\rm K} \equiv \langle \rho\bm{u}^2/2 \rangle_{\rm v}$, 
$\epsilon_{\rm Mt} \equiv \langle \bm{B}^2/2\mu_0 \rangle_{\rm v}$ and 
$\epsilon_{\rm Mm} \equiv (\langle B_x \rangle_{\rm v}^2 + \langle B_y \rangle_{\rm v}^2)/2\mu_0$ (the energy of mean magnetic 
components) are shown by solid, dashed and dash-dotted lines in Figure 2a. The evolution of the $\epsilon_{\rm K}$ of the progenitor run 
before starting the dynamo run is also shown. Note that, from the horizontal symmetry and ${\rm div} \bm{B} = 0$, 
$\langle B_z\rangle_{\rm h}$ and $\langle B_z\rangle_{\rm v}$ are zero independent of time. 

After a short relaxation time, the convective kinetic energy reaches a quasi-steady state at $t \simeq 50\tau_{\rm cv}$. The magnetic 
energy is gradually amplified by the convection and is saturated at $t \simeq 120\tau_{\rm cv}$. The mean values evaluated there are 
$u_{\rm cv} = 5.9\times 10^{-3}$ and $B_{\rm eq,v} \equiv \sqrt{\langle\langle \mu_0 \rho {\bm u}^2 \rangle\rangle_{\rm v}} 
= 7.2\times 10^{-2}$, providing $\tau_{\rm cv} = 54$ and ${\rm Ro} = 0.02$. 
The vertical profile of $\langle\langle u_z^2 \rangle\rangle_{\rm h}^{1/2}$ (solid) and the distributions of the $u_z$ on the horizontal 
planes in the dynamo-saturated stage are also shown in Figure 1b and 1c. Since the rotation gives rise to the Coriolis force 
acting on the convective motion, the convective cell shrinks and thus the scale separation becomes larger in the rotating system. 
Since the mean kinetic helicity, which is a prerequisite for exciting the large-scale dynamo, arises as a natural consequence 
of the rotation, the dynamo-generated magnetic field also affects the convective motion shown in Figure 1c.

Figure 2b shows the time-depth diagrams of $\langle B_x \rangle_{\rm h}$ and $\langle B_y \rangle_{\rm h}$ normalized by 
$B_{\rm eq} (z)$. Note that the turbulent magnetic component is eliminated by taking horizontal average. It is found that the oscillatory large-scale 
horizontal magnetic component is spontaneously organized in the bulk of the CZ. It has a peak with the super-equipartition strength in the mid-part 
of the CZ and propagates from there to the top and base of the CZ. Since there exists a phase difference of $\pi/2$ between 
$\langle B_x \rangle_{\rm h}$ and $\langle B_y \rangle_{\rm h}$, the mean horizontal magnetic flux, defined by 
$B_h \equiv \sqrt{\langle B_x \rangle_{\rm h}^2 + \langle B_y \rangle_{\rm h}^2}$, has a quasi-steady vertical profile. 

The large-scale dynamo observed here in the strongly-stratified model has physical properties similar to those in the weakly-stratified 
convective dynamo simulations \citep[e.g.,][MS14a]{kapyla+13}. Because of the horizontal symmetry and thus no differential rotation 
in our system, the turbulent electro-motive force would be solely responsible for the dynamo (see MS14b for a mean-field $\alpha^2$-dynamo 
model which can quantitatively reproduce the DNS results). Our intriguing finding in this Letter, which have not been observed in the 
weakly-stratified model with the similar boundary conditions, is spontaneous formation of large-scale magnetic structures 
in the CZ surface, which will be reported in the following. 
\subsection{Spontaneous Formation of Surface Magnetic Structure}
\begin{figure*}[tbp]
\begin{center}
\includegraphics[width=17cm,clip]{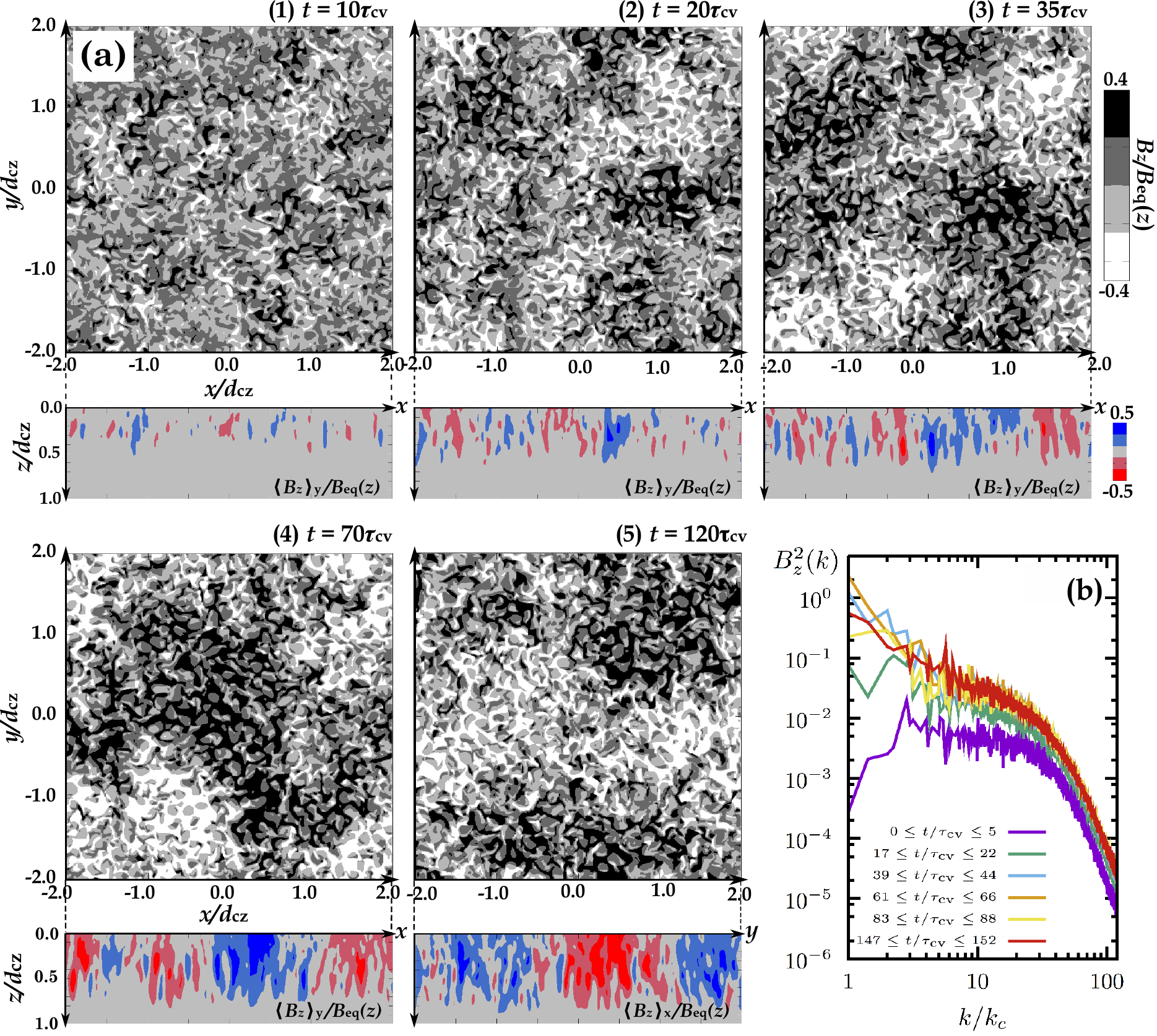}
\caption{(a) A series of snapshots for the horizontal distribution of the $B_z$ at $z/d_{\rm cz} = 0.04$ and vertical distribution of the 
$\langle B_z \rangle_{\rm y}/B_{\rm eq}(z)$ or $\langle B_z \rangle_x/B_{\rm eq}(z)$. (b) The temporal evolution of 2D Fourier 
spectrum of the $B_z^2$ in the upper CZ, where $k_c = 2\pi/W$.} 
\label{fig3}
\end{center}
\end{figure*}
A series of snapshots where the distribution of the $B_z$ at different time on the horizontal cutting plane at $z/d_{\rm cz} = 0.04$ is shown 
in the top panel of Figure 3a. The darker (lighter) tone denotes positive (negative) $B_z$. While the $B_z$ has a small-scale tangled structure 
with the typical size comparable to the convective cell in the initial evolutionary stage [(a1)--(a2)], it evolves as time passes to organize the large-scale 
structure with the spatial-scale much larger than the convective cell [(a3)--(a4)]. The surface magnetic structure has the dynamically-important 
strength comparable to $B_{\rm eq}(z)$ and is recurrently-appeared in the dynamo-saturated stage [(a5)], implying that it should be a  
consequence of the strongly-stratified MHD convection rather than a transiently-formed structure. 

The one-sided horizontal average ($x$- or $y$-average) of the $B_z$ may be helpful to clearly demonstrate its 
depth-dependence because the horizontal average of it, i.e., $\langle B_z \rangle_{\rm h}$, is zero in our setup. Shown in 
lower panel of Figure 3a is a corresponding time series of snapshots for the distributions of the $\langle B_z \rangle_{\rm y}$ 
[(a1)--(a4)] or $\langle B_z \rangle_{\rm x}$ [(a5)] in units of $B_{\rm eq} (z)$. The blue (red) tone represents the positive (negative) value. 
The initial small-scale vertical magnetic structure gradually evolves to the larger scale in the upper CZ. While it prevails in the mid-part of the 
CZ in the dynamo-saturated stage, the core of the large-scale $B_z$ structure stays in around the CZ surface. 

Figure 3b shows the temporal evolution of two-dimensional (2D) Fourier spectrum of the $B_z^2$ in the upper CZ. Here the spectrum at each 
depth is projected onto a one-dimensional wavenumber $k^2 = k_x^2 + k_y^2$ and then is averaged over the normalized depth from $0.0$ 
to $0.25$. The energy-containing scale of the $B_z$ becomes larger with the time and finally reaches the possible largest scale in the 
upper CZ, confirming the existence of the large-scale $B_z$ structure in the upper CZ. Since the large-scale vertical magnetic 
component cannot be generated solely by the $\alpha^2$-type dynamo only with the vertical helicity variation (see, MS14b), some additional 
mechanism should play a role in organizing it in this simulation.
\subsection{What Makes Surface Magnetic Structure Organized ?}
\begin{figure*}[tbp]
\begin{center}
\begin{minipage}{0.45\hsize}
\begin{center}
\includegraphics[width=7.5cm,clip]{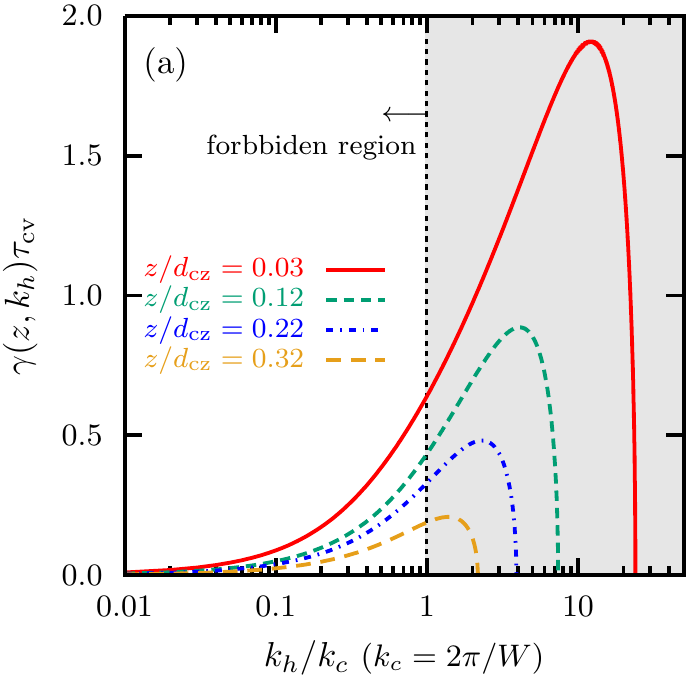}
\end{center}
\end{minipage}
\begin{minipage}{0.45\hsize}
\begin{center}
\includegraphics[width=7.9cm,clip]{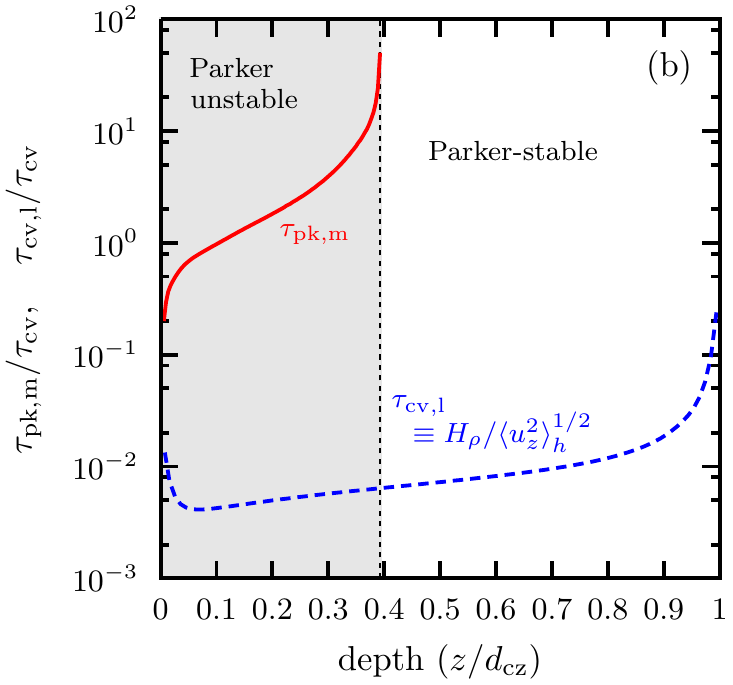}
\end{center}
\end{minipage}
\caption{(a) The growth rate of the Parker instability evaluated from the simulation result. (b) The comparison between the growth time of 
the fastest-growing mode of the Parker instability ($\tau_{\rm pk,m}$) and the local convective turn-over time ($\tau_{\rm cv,l}$) 
as a function of the depth.}
\label{fig5}
\end{center}
\end{figure*}
The so-called negative magnetic pressure instability (NEMPI) has been proposed as a mechanism for the self-assembly of 
the magnetic flux near the CZ surface \citep[e.g.,][]{kleeorin+96,brandenburg+10}. Although its presence has been confirmed numerically 
in the forced MHD turbulence \citep[][]{brandenburg+11,brandenburg+13,mitra+14,jabbari+14,warnecke+13}, it would not play a 
significant role in organizing the surface magnetic structure in our simulation. This is because the relatively-rapid rotation with ${\rm Ro} =0.02$ 
is assumed here, but according to \citet{lasoda+12}, ${\rm Ro} \gtrsim 5 $ is required for exciting the NEMPI. Here possible two alternatives 
are discussed as a cause of the magnetic structure formation: one is the Parker instability \citep[][]{parker79} and the other is the flux 
expulsion accompanied with the strong downflow \citep[e.g.,][]{weiss66, kitiashvili+10, stein+12, kapyla+15}. 

Figure 4a shows the growth rate of the Parker instability obtained from the WKB dispersion equation  
\begin{eqnarray}
(c_\star^2 + v_{A}^2 )\gamma^4 & + & v_{A}^2\left[ 2(c_\star^2 + v_A^2)k_h^2 + g_0 \frac{D}{Dz}\ln\left(\frac{B_h}{\rho}\right)\right] \gamma^2  \\ \nonumber
& + & k_h^2v_A^4\left( k_h^2 c_\star^2 + g_0\frac{D}{Dz}\ln B_h \right)  = 0 \;,
\end{eqnarray}
\citep[e.g.,][]{gilman70}, where $\gamma$ is the growth rate, $k_h$ is the horizontal wavenumber, $B_h$ is the horizontal magnetic flux, 
$c_\star \equiv \sqrt{P/\rho} $, and $v_{A} \equiv B_h/\sqrt{\mu_0 \rho}$. For deriving the depth-dependent growth rate, we adopt 
$P = \langle P \rangle_{\rm h}$, $\rho = \langle \rho \rangle_{\rm h}$ and $B_h = \sqrt{\langle B_x \rangle_{\rm h}^2 + \langle B_y \rangle_{\rm h}^2}$ 
evaluated from the simulation model. The different line-type denotes the growth rate at the different depth. The vertical and horizontal axes are 
normalized by $1/\tau_{\rm cv}$ and $k_c = 2\pi/W$. Note that the instability is inhibited in the range $k/k_c < 1$ due to the box-width constraint. 

The dynamo-maintained magnetic flux is unstable to the Parker instability in the span $0 \le z/d_{\rm cz} \lesssim 0.4$. Since the 
typical growth time of it is comparable to or a bit smaller than $\tau_{\rm cv}$, it has a sufficient time to grow during the simulation.  
However, the most unstable mode has a smaller wavelength in comparison with the box-width, implying the difficulty to explain the dominance of the 
box-sized surface magnetic structure in our simulation. 

Additionally to the mismatch of the magnetic spatial-scales, the growth of the Parker instability itself may be inhibited by vigorous 
convective motions. In Figure 4b, the growth time of the most unstable mode of the Parker instability (solid: $\tau_{\rm pk,max}$) and the local 
convective turn-over time defined by $\tau_{\rm cv, l} (z) = H_{\rm \rho} (z)/\langle u_{z}^2 \rangle_{\rm h}^{1/2}$ (dashed) are compared as a 
function of the depth, where $H_{\rho} \equiv {\rm d}z/{\rm d}\ln\langle\rho\rangle_{\rm h}$.  In the upper CZ, the condition 
$\tau_{\rm cv, l} \ll \tau_{\rm pk,max}$ is always satisfied. Since, in such a situation, the small-scale convective motion violently disturbs the 
coherency of the magnetic flux, we would have to say that the Parker instability would not be responsible for the large-scale structure 
formation observed in our simulation.

Next, the large-scale flow and its association with the surface magnetic structure are analyzed. For casting light on the 
large-scale pattern, the small-scale structures with $k/k_c \gtrsim 8$ are eliminated by applying Fourier filtering \citep[e.g.,][]{warnecke+15,jabbari+16}. 
A series of snapshots where $\langle B_z \rangle_{k_8}$ and $\langle u_z \rangle_{k_8}$ on the horizontal plane at $z/d_{\rm cz} = 0.04$ are shown in 
Figure 5a, where $\langle \cdot \rangle_{k_8}$ denotes Fourier filtering. The over-plotted arrows are the velocity vectors composed of 
$\langle u_x \rangle_{k_8}$ and $\langle u_y \rangle_{k_8}$. Additionally, 2D spectra of $B_h^2$, $B_z^2$, $\rho v_h^2$ and $\rho v_z^2$ 
in the upper CZ are also shown in Figure 5b. The spectrum at each depth is spatially-averaged over the normalized depth from $0.0$ to 
$0.25$ and is temporally-averaged over $10\tau_{\rm cv}$ around the corresponding reference time. 

It is prominent that, in the dynamo-saturated stage, the bipolar ``band-like" structure elongated along the direction of the 
horizontal magnetic flux is predominant (see Fig.2b). Although the faster horizontal flow and stronger downflow can be found in/around the 
region with the stronger $B_z$ before the dynamo-saturation (left column), large-scale flow pattern is not necessarily associated with the 
magnetic structure in the dynamo-saturated stage (middle and right columns). In addition, we can find from the spectra that the energy 
contained in the large-scale magnetic components is much larger than that of the large-scale flow in the upper CZ. This would suggest 
that the large-scale flows are not the cause, but a consequence of the large-scale magnetic structures in the upper CZ.

Overall our analyses indicate that there should be an as-yet-unknown mechanism for the self-organization of large-scale 
magnetic structures, which would be inherent in the strongly-stratified atmosphere. The band-like magnetic structure observed in our 
simulation is similar to that observed in the large-scale dynamo by the forced turbulence in a strongly-stratified atmosphere 
\citep[][]{mitra+14,jabbari+16}. This may imply that the surface magnetic structure formation is a common universal 
feature of the strongly-stratified MHD turbulence regardless of its details. 
\begin{figure*}[tbp]
\begin{center}
\includegraphics[width=18cm,clip]{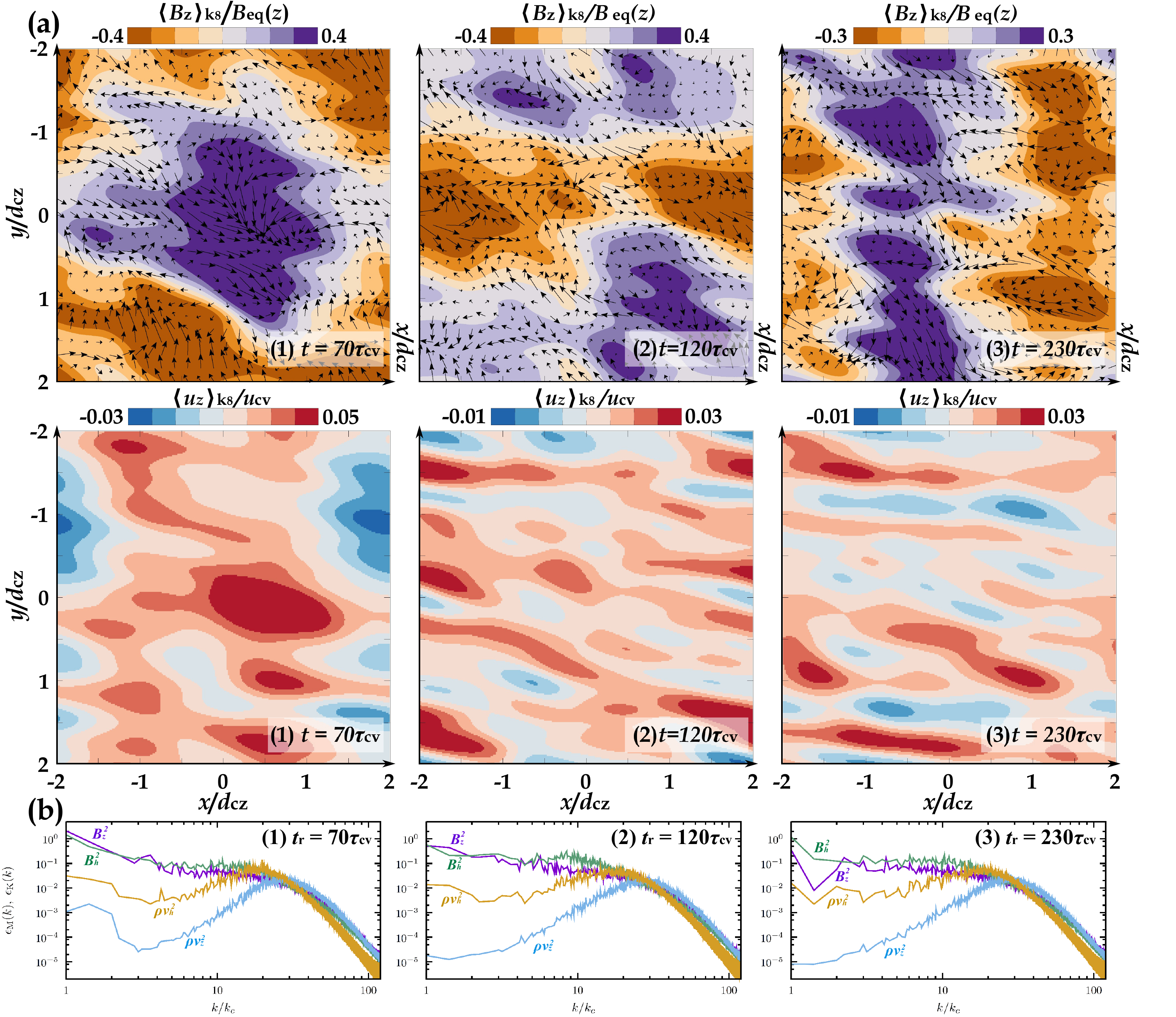}
\caption{
(a) A series of snapshots of $\langle B_z \rangle_{k_8}$ and $\langle u_z \rangle_{k_8}$ on the horizontal plane at $z/d_{\rm cz} = 0.04$. 
The over-plotted arrows are the velocity vectors. (b) 2D spectra of $B_h^2$, $B_z^2$, $\rho v_h^2$ and $\rho v_z^2$ in the upper CZ 
at around the reference time $t_r$. } 
\label{fig5}
\end{center}
\end{figure*}
\section{Summary}
In this Letter, we studied numerically MHD convections in the strongly-stratified atmosphere resembling the solar interior. 
The large-scale dynamo observed in our simulation had physical properties similar to those observed in the weakly-stratified model (see 
MS14ab): the oscillatory large-scale horizontal magnetic component with the dynamically-important strength was spontaneously organized 
in the bulk of the CZ. Its spatio-temporal evolution strongly suggests that the $\alpha^2$-type mechanism would be responsible for the MHD 
dynamo in our system.

Our intriguing finding, which have not been observed in the weakly-stratified model, was the spontaneous formation of the large-scale 
$B_z$ structure in the CZ surface. Small-scale tangled components of the $B_z$ which was seen in the earlier evolutionary stage gradually 
evolved to the large-scale organized structure with the size much larger than convective cell as time passes in the upper CZ. In the 
dynamo-saturated stage, the bipolar ``band-like" $B_z$ structure was predominant and was recurrently-appeared. Since the possible 
candidates, such as the NEMPI, Parker instability, and flux expulsion due to the strong downflow, had difficulties to explain the surface 
magnetic structure formation, our results may suggest the existence of an as-yet-unrecognized mechanism inducing the spontaneous 
formation of the large-scale magnetic structure in the strongly-stratified convective atmosphere.

The large-scale dynamo observed here maintains the quasi-steady magnetic flux and thus is different from the actual solar dynamo with a 
quasi-periodic flux modulation. Furthermore, since we adopt the faster rotation than that achieved in the actual Sun and ignore the global 
effects, such as differential rotation and meridional flows, our model still remains a long way from the solar interior.  However, despite a lack 
of some solar elements, it would be interesting that the relatively shallow root of the surface magnetic structure, implied from our simulation, 
seems to be compatible with some observations of the magnetic patches on the solar surface \citep[e.g.,][]{brandenburg05,hara09}. 
Understanding the self-organization mechanism of the surface magnetic structure observed in our study should deepen our knowledge 
of the solar magnetism and is a priority target of our future work. 
\acknowledgments
We acknowledge an anonymous referee for constructive comments. Computations were carried out on XC30 at NAOJ. This work 
was supported by JSPS KAKENHI Grant number 15K17611 and the joint research project of the Institute of Laser Engineering, 
Osaka University.


\begin{thebibliography}{32}
\expandafter\ifx\csname natexlab\endcsname\relax\def\natexlab#1{#1}\fi
\bibitem[Augustson et al.(2015)]{augustson+15} Augustson, K., Brun, A.S., Miesch, M., \& Toomre, J.\ 2015, \apj, 809, 149 
\bibitem[Brandenburg(2005)]{brandenburg05} Brandenburg, A.\ 2005, \apj, 625, 539
\bibitem[Brandenburg et al.(2010)]{brandenburg+10} Brandenburg, A., Kleeorin, N., \& Rogachevskii, I.\ 2010, Astronomische Nachrichten, 331, 5
\bibitem[Brandenburg et al.(2011)]{brandenburg+11} Brandenburg, A., Kemel, K., Kleeorin, N., Mitra, D., \& Rogachevskii, I.\ 2011, \apjl, 740, L50 
\bibitem[Brandenburg et al.(2013)]{brandenburg+13} Brandenburg, A., Kleeorin, N., \& Rogachevskii, I.\ 2013, \apjl, 776, L23 
\bibitem[Charbonneau(2010)]{charbonneau10} Charbonneau, P.\ 2010, Living Reviews in Solar Physics, 7, 3 
\bibitem[Cheung et al.(2010)]{cheung+10} Cheung, M.C.M., Rempel, M., Title, A.M., \& Sch{\"u}ssler, M.\ 2010, \apj, 720, 233 
\bibitem[Christensen-Dalsgaard et al.(1996)]{CD+96} Christensen-Dalsgaard, J., et al.\ 1996, Science, 272, 1286 
\bibitem[{{Clarke}(1996)}]{clarke96}{Clarke}, D.A. 1996, \apj, 457, 291
\bibitem[Evans \& Hawley(1988)]{evans+88} Evans, C.R., \& Hawley, J.F.\ 1988, \apj, 332, 659 
\bibitem[Fan \& Fang(2014)]{fan+14} Fan, Y., \& Fang, F.\ 2014, \apj, 789, 35
\bibitem[Ghizaru et al.(2010)]{ghizaru+10} Ghizaru, M., Charbonneau, P., \& Smolarkiewicz, P.K.\ 2010, \apjl, 715, L133 
\bibitem[Gilman(1970)]{gilman70} Gilman, P.~A.\ 1970, \apj, 162, 1019 
\bibitem[Hara(2009)]{hara09} Hara, H.\ 2009, \apj, 697, 980
\bibitem[Jabbari et al.(2014)]{jabbari+14} Jabbari, S., Brandenburg, A., Losada, I.~R., Kleeorin, N., \& Rogachevskii, I.\ 2014, \aap, 568, A112 
\bibitem[Jabbari et al.(2016)]{jabbari+16} Jabbari, S., Brandenburg, A., Mitra, D., Kleeorin, N., \& Rogachevskii, I.\ 2016, arXiv:1601.08167 
\bibitem[K{\"a}pyl{\"a} et al.(2012)]{kapyla+12} K{\"a}pyl{\"a}, P.J., Mantere, M.J., \& Brandenburg, A.\ 2012, \apjl, 755, L22 
\bibitem[K{\"a}pyl{\"a} et al.(2013)]{kapyla+13} K{\"a}pyl{\"a}, P.J., Mantere, M.J., \& Brandenburg, A.\ 2013, GAFD, 107, 244 
\bibitem[K{\"a}pyl{\"a} et al.(2015)]{kapyla+15} K{\"a}pyl{\"a}, P.J., Brandenburg, A., Kleeorin, N., K{\"a}pyl{\"a}, M.J., \& Rogachevskii, I.\ 2015, arXiv:1511.03718 
\bibitem[Kitiashvili et al.(2010)]{kitiashvili+10} Kitiashvili, I.N., Kosovichev, A.G., Wray, A.A., \& Mansour, N.N.\ 2010, \apj, 719, 307 
\bibitem[Kleeorin et al.(1996)]{kleeorin+96} Kleeorin, N., Mond, M., \& Rogachevskii, I.\ 1996, \aap, 307, 293 
\bibitem[Losada et al.(2012)]{lasoda+12} Losada, I.R., Brandenburg, A., Kleeorin, N., Mitra, D., \& Rogachevskii, I.\ 2012, \aap, 548, A49 
\bibitem[Masada et al.(2013)]{masada+13} Masada, Y., Yamada, K., \& Kageyama, A.\ 2013, \apj, 778, 11 
\bibitem[Masada \& Sano(2014a)]{masada+14a} Masada, Y., \& Sano, T.\ 2014 \pasj, 66, S2 
\bibitem[Masada \& Sano(2014b)]{masada+14b} Masada, Y., \& Sano, T.\ 2014, \apjl, 794, L6 
\bibitem[Miesch(2005)]{miesch05} Miesch, M.S.\ 2005, Living Reviews in Solar Physics, 2, 1 
\bibitem[Mitra et al.(2014)]{mitra+14} Mitra, D., Brandenburg, A., Kleeorin, N., \& Rogachevskii, I.\ 2014, \mnras, 445, 761 
\bibitem[Nelson et al.(2013)]{nelson+13} Nelson, N.J., Brown, B.P., Brun, A.S., Miesch, M.S., \& Toomre, J.\ 2013, \apj, 762, 73 
\bibitem[Parker(1979)]{parker79} Parker, E. N.\ 1979, Oxford University Press, 1979, 858 p.,  
\bibitem[Rempel \& Cheung(2014)]{rempel+14} Rempel, M., \& Cheung, M.C.M.\ 2014, \apj, 785, 90 
\bibitem[{{Sano} {et~al.}(1998){Sano}, {Inutsuka}, \& {Miyama}}]{sano+98}{Sano}, T., {Inutsuka}, S., \& {Miyama}, S.~M. 1998, \apjl, 506, L57
\bibitem[Spruit et al.(1990)]{spruit+90} Spruit, H.C., Nordlund, A., \& Title, A.M.\ 1990, \araa, 28, 263 
\bibitem[Stein \& Nordlund(2012)]{stein+12} Stein, R.F., \& Nordlund, {\AA}.\ 2012, \apjl, 753, L13 
\bibitem[Warnecke et al.(2013)]{warnecke+13} Warnecke, J., Losada, I.~R., Brandenburg, A., Kleeorin, N., \& Rogachevskii, I.\ 2013, \apjl, 777, L37
\bibitem[Warnecke et al.(2015)]{warnecke+15} Warnecke, J., Losada, I.~R., Brandenburg, A., Kleeorin, N., \& Rogachevskii, I.\ 2015, arXiv:1502.03799 
\bibitem[Weiss(1966)]{weiss66} Weiss, N.O.\ 1966, Proceedings of the Royal Society of London Series A, 293, 310 
\bibitem[Yadav et al.(2015)]{yadav+15} Yadav, R.K., Gastine, T., Christensen, U.R., \& Reiners, A.\ 2015, \aap, 573, A68 
\end{thebibliography}
\end{document}